\documentclass[12pt]{iopart}
\usepackage{graphicx}
\begin{document}

\title[Mechanism of electron localization in a QW]{Mechanism of electron
localization in a quantum wire}

\author{B S Shchamkhalova, V A Sablikov}

\address{Kotel'nikov Institute of Radioengineering and Electronics,
Russian Academy of Sciences, Fryazino, Moscow District, 141190,
Russia} \ead{bagun@ire216.msk.su}
\begin{abstract}

 We show that quasi-bound electron states are formed in a
quantum wire as a result of electron backscattering in the
transition regions between the wire and the electron reservoirs,
to which the wire is coupled. The backscattering mechanism is
caused by electron density oscillations arising even in smooth
transitions due to the reflection of electrons not transmitting
through the wire. The quasi-bound states reveal themselves in
resonances of the electron transmission probability through the
wire. The calculations were carried out within the Hartree-Fock
approximation using quasiclassic wavefunctions.

\end{abstract}

\maketitle

Quantum point contacts (QPCs) and quantum wires (QWs) have
attracted much interest as model systems for studying effects of
an electron-electron interaction in one-dimensional (1D) systems.
The conductance of these devices is known~\cite{van Wees,Wharam}
to be quantized according to a universal law, $G=2ne^2/h$, where
$n=1,2,3 \dots$, which was successfully explained within the model
of noninteracting electrons~\cite{Glazman,Buttiker}. However,
recent experiments have revealed a lot of other features of the
conductance which are yet unexplained. Widely discussed is the
0.7~anomaly~\cite{Thomas}. Besides, such observations as strongly
nonlinear conductance at low applied voltage~\cite{Picciotto},
additional plateaulike features~\cite{Reilly} and even resonances
in the differential
conductance~\cite{Reilly,Morimoto,Shailos,Biercuk} are also of
interest. It is obvious that these features are caused by the
electron-electron interaction, but only electron-electron
interaction in the 1D system fails to explain them. Of no less
importance is the fact that a 1D wire is coupled to 2D electron
reservoirs and hence there are transition regions between the 1D
QW and the 2D reservoirs (1D-2D junctions). The most puzzling
result found recently is the electron localization in the QPC,
which was first brought to light from the studies of the 0.7
plateaulike feature in short QWs. Cronenwett~\emph{et
al}~\cite{Cronenwett} related this feature to the Kondo effect
caused by the electron spin localization in the QW. The most
convincing evidence of the electron localization was provided by
the momentum-resolved tunnelling experiments of
Auslaender~\emph{et al}~\cite{Auslaender}. Peaks and kinks in
conductance dependences on the gate voltage observed on the
devices with high electron mobility~\cite{Morimoto,Reilly,Shailos}
implicitly also points to the presence of quasi-bound electron
states. The problem is that in all the experiments the electrons
are localized over the barrier formed by the electrodes, the
potential of which varies smoothly on the electron wavelength
scale. The localization mechanism remains unclear, but it is
widely believed that the existence of the quasi-bound states in
the QW allows one to interpret the plateualike features and the
resonances of the conductance. In \cite{Jefferson} the quasi-bound
states were related to local broadening of the QW and formation of
a potential well. To justify the formation of the quasi-bound
states, Hirose~\emph{et al}~\cite{Hirose} calculated electron
density distribution in QPCs with geometric lengths of the order
of the Fermi wavelength. However in the experiments the geometric
sizes exceed the Fermi wavelength noticeably, the barrier induced
by the gates is rather smooth and the over-barrier reflection is
negligible. Recently, Rejec and Meir~\cite{Rejec} have
demonstrated the presence of quasi-bound states in the QPC by
calculations based on spin-density functional theory, but the
underlying physical mechanism remains unknown.

Such a mechanism was suggested in \cite{Sablikov}. The
localization is a result of an intersubband electron interaction
in the 1D-2D transition regions. The interaction is caused by the
Friedel oscillations of the density of higher subband electrons,
which do not pass through the transition regions, and are
reflected. The electrons of an open subband are backscattered by
these oscillations. Since the backscattering occurs in two opposed
sides of the QW, quasi-bound states are formed. When studying the
backscattering in 1D-2D junctions the problem is the complicated
structure of the electron density distribution along the QW.
Besides a smooth component of the electron density there is an
oscillating one, and it is just this component that gives rise to
the electron backscattering. An analytic theory of scattering by
the Friedel oscillations in the outside of the junction (the far
zone), where the wave vector of the oscillations is close to
$2k_F$ and the electrons with the Fermi energy are resonantly
scattered, was developed in \cite{Sablikov}. But the scattering by
nonperiodic oscillations of the electron density inside the
transition region (the near zone) may be also important, because
the oscillation amplitude is larger there. In this paper the
electron backscattering in the smooth 1D-2D junctions is studied
within the Hartree-Fock approximation taking into account both
near and far zones. The backscattering in the near zone turns out
to contribute essentially to the total effect . It is found that
quasi-bound states are formed in the QW giving rise to the
transmission resonances.

\begin{figure}
\centerline{\includegraphics[width=10 cm]{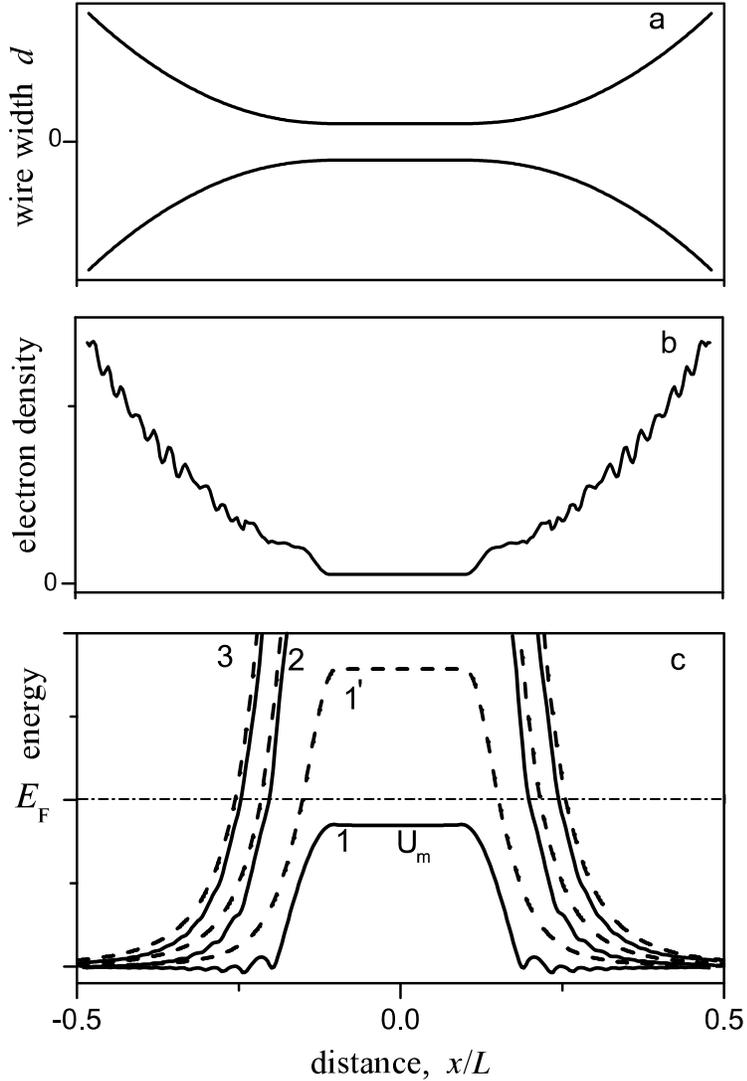}} \caption{(a)
Sketch of the device. (b)  Effective 1D electron density. (c)
Potential landscape and subband energies. Solid lines --
selfconsistent relief $U(x)$ of the first (1) and higher (2,3)
subband bottoms; dashed lines -- transverse quantization energies
of the first (1') and higher subbands.} \label{d_rho_U}
\end{figure}

Consider a QW in the form of a strip connecting 2D electron
reservoirs with a given electrochemical potential $E_F$. The strip
width, $d(x)$, varies as follows:
\begin{equation}\label{junc_form1}
d(x)=\left\{
\begin{array}{ll}
d = \mathrm {const} , & |x|<a\\
d[1+(|x|-a)^2/R^2], & |x|>a \,,
\end{array} \right.
\end{equation}
where the broadening radius $R$ considerably exceeds both $d$ and
$k_F^{-1}$ ($k_F$ is the Fermi wave vector in reservoirs). For
simplicity, we assume that only the first subband is open. The
electrons in the higher subbands are reflected in the 1D-2D
junctions.

The backscattering of electrons incident on the QW is calculated
in a way similar to that of \cite{Sablikov} with two essential
additions which are required to include the scattering process in
the near zone. First, in addition to the Friedel oscillations
produced by closed subbands we take into account the electron
density oscillations produced by the first subband electrons with
energy below the potential landscape maximum in the QW, $U_m$,
(figure~\ref{d_rho_U}(c)). These electrons are reflected from the
barrier giving rise to electron density oscillations with the wave
vector $\sim 2\sqrt{2mU_m}/\hbar$, which is close to $2k_F$.
Hence, these oscillations can noticeably contribute to the
backscattering of the electrons passing through the QW. The second
addition is the selfconsistent calculation of the smooth
components of the potential and the electron density distribution
in the QW. The potential landscape along the QW axis is formed by
the transverse quantization energy and the smooth component of the
Hartree potential. This is important because in the near zone the
potential landscape considerably differs from the transverse
quantization energy, as figure~\ref{d_rho_U}(c) shows.

The calculations are carried out in the following way. First,
one-particle wave functions are written in the adiabatic
approximation as a product of transverse and longitudinal wave
functions
$$\Psi_n(\vec{r}_\perp,x)=\phi_{nx}(\vec{r}_\perp)\psi_n(x).
$$
where $n=1,2,\dots$ is the subband number,
$\phi_{nx}(\vec{r}_\perp)$ is a transverse wavefunction
corresponding to transverse quantization energy $E_n(x)$. Second,
effectively 1D equations are obtained for the longitudinal wave
functions $\psi_n(x)$ by averaging the Hartee-Fock equations over
transverse coordinates with the weight
$\phi^*_{nx}(\vec{r}_\perp)$. As a result one obtains 1D
Schr\"odinger equations with effective 1D Hartree and exchange
terms. Third, these equations are solved selfconsistently using
perturbation theory. To zero order in the interaction, the wave
functions are written in the quasi-classic approximation. At this
stage the problem is solved numerically using the iteration
procedure developed in \cite{Shchamkhalova}. The electron
scattering is calculated in the first Bohrn approximation. The
quasi-classic approximation is justified if the local wavelength
of an electron on the Fermi level is smaller than the
characteristic spatial scale of the potential~\cite{Zeldovich}.
The adiabatic approximation and its application to similar
structures were considered in detail by Glazman et
al~\cite{Glazman,Glazman1}

Thus, zero-order wave functions for the closed states are
\begin{equation}
\label{c_band} \psi_{n,k}(x)= 2 \sqrt{\frac{k}{k_n(x)}
}\cos\left[\int^x_{a_n}\!\!dx' k_n(x') -\frac{\pi}{4}\right]\, ,
\end{equation}
where  $k_n(x)$ is the wave vector of the longitudinal motion,
$k=\lim_{x\to \infty}k_n(x)$ and $a_n(k)$ is a turning point. For
electrons of the first subband with energy higher than the
potential landscape maximum the zero-order wavefunction is
\begin{equation}
\label{o_band} \psi_{1,k}(x)= \sqrt{\frac{k}{k_1(x)}} \exp
\left[i\int^x_{0}\!\!\!dx' k_n(x') \right].
\end{equation}

\begin{figure}
\centerline{\includegraphics[width=12cm]{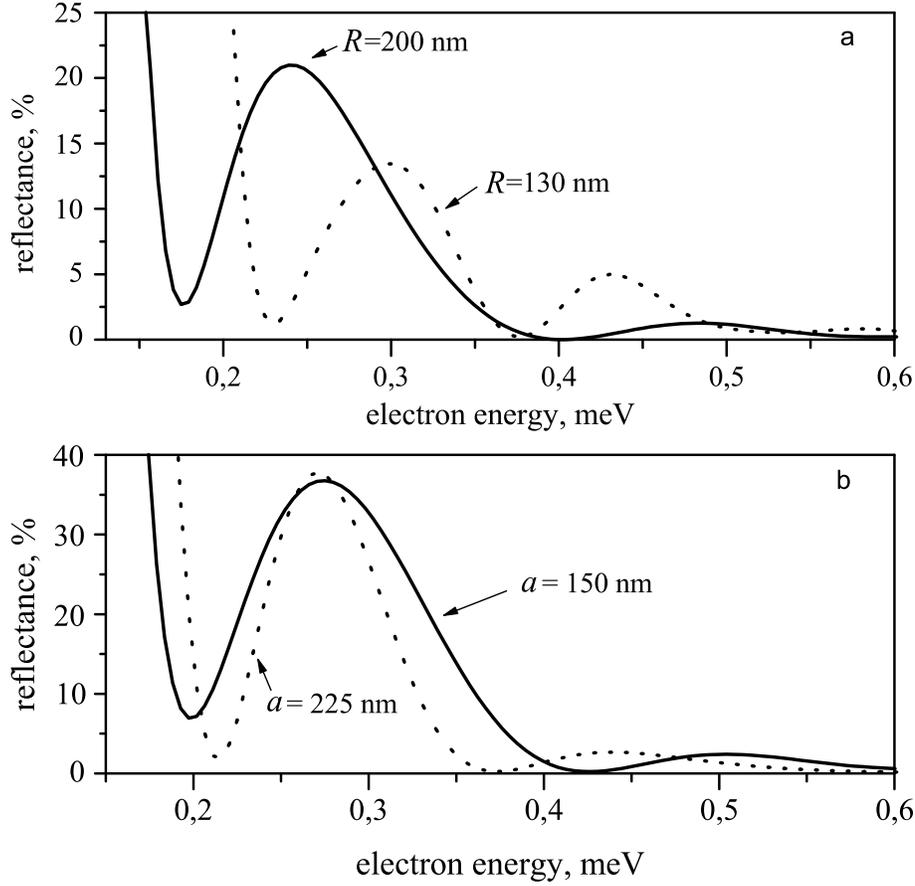}}
\caption{Reflection coefficient for different geometric sizes: (a)
the widening radius $R$ is varied at $a = 150$~nm, $D=600$\AA,
$E_F=9$~meV; (b) the length $2a$ of uniform part is varied at
$R=135$~nm, $D= 900$\AA, $E_F=9$~meV. For $a = 150$~nm (solid
line), transmission resonances at energies 0.2 and 0.42~meV are
caused by second and third quasi-bound states in the QW. For
$a=225$~nm (dotted line) the transmission resonances correspond to
third and fourth states.} \label{reflect_size}
\end{figure}

The electron density in each subband, having been found using
equations (\ref{c_band}) and (\ref{o_band}),  depends on the
electrochemical potential in the reservoirs $E_F$ and the
effective potential $U^n(x)$ in the device. The density is a sum
of two components: one is oscillating and the other varies
smoothly on the electron wavelength scale. Accordingly, the
potential acting on the electrons also has two similar components.
A smooth component of the potential is calculated selfconsistently
with the electron density using the technique developed in
\cite{Shchamkhalova}. As a result, the potential profile $U^n(x)$
(see Fig.\ref{d_rho_U}) is obtained  for each subband and used to
calculate the wave vectors $k_n(x)$ in equations (\ref{c_band})
and (\ref{o_band}), $k_n(x) = \sqrt{k^2-2mU^n(x)/h^2}$. A rapidly
oscillating component of the potential is considered as a
perturbation, which scatters an electron from a state $|n,k\rangle
$ to $|m,k'\rangle$. The scattering potential contains the Hartree
and exchange components:
$\hat{V}=V^H+\hat{V}^{exc}$~\cite{Sablikov}. The Hartree potential
acting on the first subband electrons is
$$ V^H(x) = \sum_l \int dx'
V^H_{1,l}(x,x') \rho_l(x') - \int dx' V^H_{1,0}(x,x')
\rho_0(x')\,,
$$
where $\rho_l(x)$ is the electron density in the \emph{l}th
subband, $\rho_0(x)$ is the background charge density and
$$
V_{nl}(x,x')=\int\,V(\vec r,\vec r')|\phi_{nx}(\vec{r}_\perp)|^2
|\phi_{lx}(\vec{r'}_\perp)|^2d\vec{r}_\perp d\vec{r'}_\perp,
$$
$$
V_{n,0}(x,x')=\int\,V(\vec r,\vec r')|\phi_{nx}(\vec{r}_\perp)|^2
d\vec{r}_\perp d\vec{r'}_\perp.
$$
The exchange interaction is described by an operator, which has
following form for the first subband electrons
$$ \hat{V}^{exc} \psi_{1,k} (x) = - \sum_l \int dx'
V^{exc}_{1,l}(x,x') \rho_l(x,x') \psi_{1,k}(x') \, ,
$$
where $\rho_l(x,x')$ is the density matrix and
\begin{eqnarray*}
V^{exc}_{n,m}(x,x')=\int\!\!\! d\vec{r}_{\perp}
d\vec{r'}_{\perp} V({\bf r},{\bf r'}) 
\phi_{nx}(\vec{r}_\perp) \phi_{mx}(\vec{r}_\perp)
\phi_{nx}(\vec{r'}_\perp) \phi_{mx}(\vec{r'}_\perp)\, .
\end{eqnarray*}
Here $V({\bf r},{\bf r'})$ is the pair interaction potential. The
screening of the Coulomb interaction is taken into account
similarly to \cite{Shchamkhalova}, assuming that the screening is
produced by a conducting plane (gate) situated over the device at
a distance $D$ or/and by the reservoirs, deep inside which the
potential is fixed. In specific calculations the potential is
taken to be constant at a distance $\pm L/2$ from the center of
the QW. The reflection amplitude for electrons in the open subband
(i.e.,for the $(1,k) \to (1,-k)$ transition) is calculated in the
Born approximation
$$
r_k=\frac{m}{i\hbar^2}\int \!dx\, \psi^*_{1,k} \hat{V}
\psi_{1,k}\, .
$$

\begin{figure}
\centerline{\includegraphics[width=12cm]{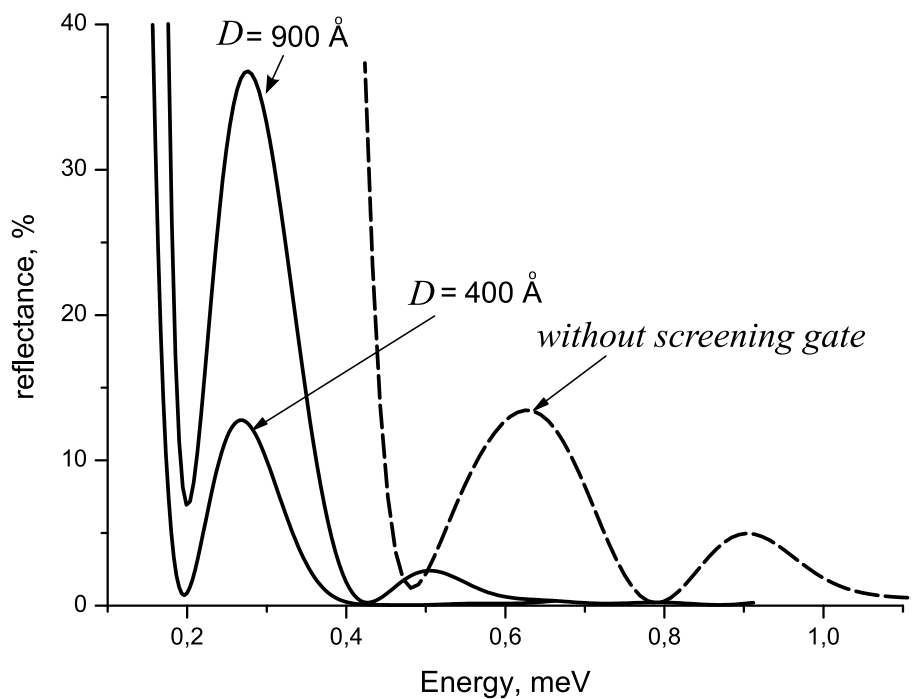}}
\caption{Reflection coefficient for different distance to the
screening gate $D$. The parameters used are $a=150$ nm; $ R=135$
nm; $L=1.5$ $\mu$m; $E_F = 9$ meV.} \label{reflect_screen}
\end{figure}

\begin{figure}
\centerline{\includegraphics[width=12 cm]{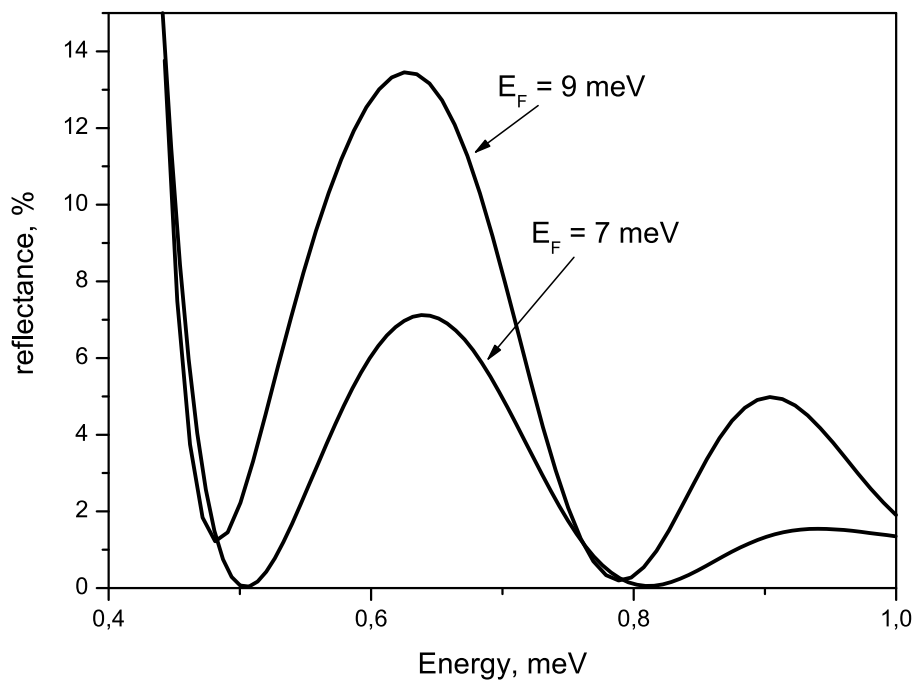}}
\caption{Reflection coefficient for different Fermi energies $E_F$
in reservoirs. The parameters used are $a=150$ nm; $ R=135$ nm;
$L=1.5$ $\mu$m; screening gate is absent.} \label{reflect_EF}
\end{figure}

The main results obtained are shown in
figures~\ref{reflect_size},\ref{reflect_screen},\ref{reflect_EF}
where the reflection coefficient $|r|^2$ of the electrons incident
on the QW is shown as a function of the electron energy measured
from the potential landscape maximum for a variety of the device
parameters. One sees the resonant behavior of the electron
reflection. At some energies the reflectance strongly diminishes
and correspondingly the transmission resonantly increases.
Calculations show that the resonance energies are mainly
determined by the geometric sizes of the device: the length of the
uniform part $2a$ and the broadening radius $R$. This is
demonstrated in figure~\ref{reflect_size}. The screening effect on
the reflectance is demonstrated by figure~\ref{reflect_screen}
where the reflectance spectra are shown for a variety of the
distances $D$ between the QW and the screening electrode. This
distance affects both the width of the transmission resonances
(the width of the resonance decreases with $D$) and the reflection
coefficient at energies between the resonances ($|r|^2$ increases
with $D$), the resonance energies being weakly dependent on $D$.
Similarly, the Fermi level in the reservoirs weakly affects the
position of the resonance, while $|r|^2$ is affected
noticeably(figure~\ref{reflect_EF}).

The reflection resonances clearly point to the presence of
quasi-bound states located in the region of the potential
landscape maximum. The spectrum of the quasi-bound states and the
energy dependence of the backscattering may be described rather
well by a simple model. The electron scattering in 1D-2D junctions
may be imagined as scattering by two $\delta$-like barriers
located symmetrically at a distance $l$ from the QW center. The
scattering potential is $W(x)=\Omega \delta(x\pm l)$. Here $l$ and
$\Omega$ are fitting parameters. The wave vector $K$, for which
the backscattering vanishes, is defined by the equation
$$
\tan(2Kl) = -\frac{2K\hbar^2}{2m\Omega}\,.
$$
Using the ratio of energies of the sequential resonances in our
numerical results we can define the serial numbers of the
resonances. Then choosing the distance $l$ properly we can fit the
resonance energies. The fitting leads to a simple equation:
$$
2l = 2a+\gamma R,
$$
where $\gamma$ is a parameter ($\gamma \simeq 0.5$),  which only
slightly depends on the device geometry and the Fermi level in the
reservoirs. The variation of the distance $D$ and the background
positive charge density affect the power $\Omega$ of the effective
scattering potential, which affects on the resonance energies
slightly.

Thus, we have shown that the interaction between the electrons of
the different subbands in 1D-2D junctions essentially affects the
electron transport in QPCs and QWs. This interaction results in
the transmission resonances which clearly evidence the formation
of quasi-bound states in the region of the potential landscape
maximum.

\section*{Acknowledgments} The work was supported by the Russian Foundation for Basic
Research (project no. 05-02-16854) and by Russian Academy of
Sciences (programs "Quantum Nanostructures" and "Strongly
Correlated Electrons in Semiconductors, Metals, Superconductors,
and Magnetic Materials"), and the Ministry of Education and
Science of RF.

\end{document}